\documentclass[twocolumn,preprintnumbers,prb,floatfix,superscriptaddress,amsmath,amssymb]{revtex4}
\usepackage{epsfig}
\usepackage{graphicx}
\usepackage{subfigure}
\usepackage{amsmath}
\usepackage{color}
\usepackage{verbatim} 
\usepackage{amssymb}
\usepackage{graphicx}
\usepackage{dcolumn}
\usepackage{bm}
\input epsf

\def\be{\begin{equation}}
\def\ee{\end{equation}}

\def\ba{\begin{align}}
\def\eal{\end{align}}

\begin{document}

\title{Effects of network topology, transmission delays, and refractoriness on the response of coupled excitable systems to a stochastic stimulus}

\author{Daniel B. Larremore}
\email{daniel.larremore@colorado.edu}
\affiliation{Department of Applied Mathematics, University of Colorado, Boulder, CO 80309, USA}

\author{Woodrow L. Shew}
\affiliation{National Institute of Health, National Institute of Mental Health, Bethesda, MD 20892, USA}

\author{Edward Ott}
\affiliation{Institute for Research in Electronics and Applied Physics, University of Maryland, College Park, Maryland 20742,
USA}

\author{Juan G. Restrepo}
\affiliation{Department of Applied Mathematics, University of Colorado, Boulder, CO 80309, USA}

\date{\today}

\begin{abstract}
We study the effects of network topology on the response of networks of coupled discrete excitable systems to an external stochastic stimulus. We extend recent results that characterize the response in terms of spectral properties of the adjacency matrix by allowing distributions in the transmission delays and in the number of refractory states, and by developing a nonperturbative approximation to the steady state network response. We confirm our theoretical results with numerical simulations.  We find that the steady state response amplitude is inversely proportional to the duration of refractoriness, which reduces the maximum attainable dynamic range.  We also find that transmission delays alter the time required to reach steady state.  Importantly, neither delays nor refractoriness impact the general prediction that criticality and maximum dynamic range occur when the largest eigenvalue of the adjacency matrix is unity. \end{abstract}


\maketitle

{\bf 
Networks of coupled excitable systems describe many engineering, biological and social applications. Recent studies of how such networks respond to an external, stochastic stimulus have provided insight on information processing in sensory neural networks \cite{kinouchiCopelli,Larremore}.  In agreement with recent experiments\cite{woody}, these studies showed that the dynamic range of neural tissue is maximized in the critical regime, which is precisely balanced between growth and decay of propagating excitation. This regime was studied theoretically for directed random Erd\H{o}s-R\'{e}nyi networks in Ref.\cite{kinouchiCopelli}, where it was found to be characterized by a network mean degree equal to one.  However, other studies\cite{copelliCampos,wu} showed that this condition does not specify criticality for other network topologies.  In this paper, extending recent results, we present a general framework for studying the effects of network topology on the response to a stochastic stimulus.  With this framework, we  derive a requirement for criticality and maximum dynamic range that holds for a wide variety of network topologies.  Moreover, we show that this prediction holds when refractory states and transmission time delays are included in the network dynamics, although other aspects of the response do depend on these properties.}
\section{Introduction}

Many applications involve networks of coupled excitable systems. Two prominent examples are the spread of information through neural networks and the spread of disease through human populations. The collective dynamics of such systems often defy naive expectations based on the dynamics of their individual components. For example, the collective response of a neural network can encode sensory stimuli which span more than 10 orders of magnitude in intensity, while the response of a single neuron typically encodes a much smaller range of stimulus intensities. Likewise, the collective properties of social contact networks determine when a disease becomes an epidemic.

Recently, a framework to study the response of a network of coupled excitable systems to a stochastic stimulus of varying strength has been proposed. The Kinouchi-Copelli model\cite{kinouchiCopelli} considers the response of a directed Erd\H{o}s-R\'{e}nyi random network of coupled discrete excitable systems to a stochastic external stimulus. A mean-field analysis of this model predicted\cite{kinouchiCopelli} that the maximum {\it dynamic range} (the range of stimuli over which the network's response varies signiÞcantly) occurs in the critical regime where an excited neuron excites, on average, one other neuron. This criterion can be stated as the mean out-degree of the network being one, $\langle d^{out} \rangle = 1$, where the out-degree of a node $d^{out}$ is defined as the expected number of nodes an excited node excites in the next time step (Ref.\cite{kinouchiCopelli} refers to this quantity as the branching ratio). 

Subsequent studies explored this system on networks with power-law degree distributions and hypercubic lattice coupling, and with a varying number of loops \cite{Larremore,copelliCampos,ribeiroCopelli,assisCopelli, wu}, showing that the criterion for criticality based on the network mean degree does not hold for networks with a heterogeneous degree distribution. However, these studies (except\cite{Larremore}) do not take into account features that are commonly found in real networks, such as, for example, community structure, correlations between in- and out-degree of a given node, or correlations between the degree of two nodes at the ends of a given edge \cite{newmanAssortativity}. Furthermore, they do not consider the effect of transmission delays or a distribution in the number of refractory states.

In a recent report \cite{Larremore}, we presented an analysis of the Kinouchi-Copelli model that accounts for a complex network topology. We found that the general criterion for criticality is that the largest eigenvalue of the network adjacency matrix is one, $\lambda=1$, rather than $\langle d^{out} \rangle = 1$. While this improved criterion successfully takes into account various structural properties of networks, our analysis did not address the effect of delays or multiple refractory states, and was based on perturbative approximations to the network response. In this paper, we will extend the results of Ref.\cite{Larremore} by developing a nonperturbative analysis that accounts for distributions in the transmission delays and number of refractory states.

This paper is organized as follows. In Section~\ref{background} we describe previous related work and the standard Kinouchi-Copelli model. In Section~\ref{longtitle} we present the model to be analyzed and derive a governing equation for its dynamics. In Section~\ref{analysis} we present our main theoretical results. In Section~\ref{dynamicrange} we apply our results to estimate the dynamic range of excitable networks. In Section~\ref{numerical} we present numerical experiments to validate our results. We discuss our results in Sec.~\ref{discussion}.

\section{Background}\label{background}

In this Section we describe the Kinouchi-Copelli model \cite{kinouchiCopelli} and other relevant previous work. In order to focus on the effects of network topology, the dynamics of the excitable systems is taken to be as simple as possible. The model considers $N$ coupled excitable elements. Each element $i$ can be in one of $m+1$ states, $x_{i}$. The state $x_{i}=0$ is the resting state, $x_{i}=1$ is the excited state, and there may be additional refractory states $x_{i}=2,3,...,m$. At discrete times $t=0,1,...$ the states of the elements $x_{i}^{t}$ are updated as follows: (i) If element $i$ is in the resting state, $x_{i}^{t}=0$, it can be excited by another excited element $j$, $x_{j}^{t}=1$, with probability $A_{ij}$, or independently by an external process with probability $\eta$; (ii) the elements that are excited or in a refractory state, $x_{i}^{t} \geq 1$, will deterministically make a transition to the next refractory state if one is available, or return to the resting state otherwise (i.e., $x_{i}^{t+1} = x_{i}^{t} + 1$ if $1 \leq x_{i}^{t} < m$, and $x_{i}^{t+1} = 0$ if $x_{i}^{t} = m$). 

For a given value of the external stimulation probability $\eta$, which is interpreted as the stimulus strength, the network response $F$ is defined in Ref.\cite{kinouchiCopelli} as
\begin{align}\label{response} 
F = \langle f \rangle_{t},
\end{align}
where $\langle \cdot \rangle_{t}$ denotes an average over time and $f^{t}$ is the fraction of excited nodes at time $t$. Of interest is the dependence of the response $F(\eta)$ on the topology of the network encoded by the connection probabilities $A_{ij}$. In particular, it is found that, depending on the network $A$, the network response can be of three types\cite{kinouchiCopelli,Larremore}: \textit{quiescent}, in which the network activity is zero for vanishing stimulus, $\displaystyle \lim_{\eta \to 0} F = 0$; \textit{active}, in which there is self-sustained activity for vanishing stimulus, $\displaystyle \lim_{\eta \to 0} F > 0$; and \textit{critical}, in which the response is still zero for vanishing stimulus but is characterized by sporadic long lasting avalanches of activity that cause a much slower decay in the response, compared with the quiescent case case, as the stimulus is decreased. Recent experiments \cite{woody} suggest that cultured and acute cortical slices operate naturally in the critical regime. Therefore, the network properties that characterize this regime are of particular importance.


In Ref. \cite{kinouchiCopelli}, the response $F$ was theoretically analyzed as a function of the external stimulation probability, $\eta$, using a mean-field approximation in which connection strengths were considered uniform, $A_{ij}=\langle d \rangle/N$ for all $i, j$. It was shown that the critical regime is achieved at the value $\langle d \rangle=1$, with the network being quiescent (active) if $\langle d \rangle < 1$ ($\langle d \rangle > 1$). For more general networks (i.e., $A_{ij}$ not constant), $\langle d \rangle$ is defined as the mean degree $\langle d \rangle = \frac{1}{N}\sum_{i,j} A_{ij} = \langle d^{in} \rangle = \langle d^{out} \rangle$, where $d^{in}_{i} = \sum_{j} A_{ij}$ and $d^{out}_{i} = \sum_{j} A_{ji}$ are the in- and out-degrees of node $i$, respectively, and $\langle \cdot \rangle$ is an average over nodes. Such critical branching processes result in avalanches of excitation with power-law distributed sizes. Cascades of neural activity with power-law size and duration distributions have been observed in brain tissue cultures \cite{beggsPlenz,woody,woody2011,pasquale2008,tetzlaff2010}, awake monkeys \cite{awakeMonkeys,woody2011}, and anesthetized rats \cite{anesthetizedRats,ribeiro2010,hahn2010}.  While $\langle d \rangle=1$ successfully predicts the critical regime for  Erd\H{o}s-R\'{e}nyi random networks \cite{kinouchiCopelli}, it does not result in criticality in networks with a more heterogeneous degree distribution \cite{wu,copelliCampos}. Perhaps more importantly, previous theoretical analyses \cite{kinouchiCopelli,wu,copelliCampos} are not able to take into account features that are commonly found in real networks, such as, for example, community structure, correlations between in- and out-degree of a given node, or correlations between the degree of two nodes at the ends of a given edge \cite{newmanAssortativity}. We will generalize the mean-field criterion $\langle d \rangle=1$ to account for complex interaction topologies encoded in the matrix $A$ as well as refractoriness and transmission delays.

\section{Generalized Kinouchi-Copelli model}\label{longtitle}

\subsection{Description of the model}

We will analyze a generalized version of the Kinouchi-Copelli model which includes possibly heterogeneous distributions of delays and refractory periods. The model is as follows:
\begin{itemize}
\item There are $N$ excitable elements, labeled $i = 1,\dots, N$.
\item At discrete times $t=0,1,...$, each element $i$ can be in one of $m_i+1$ states, $x_{i}^t$. The state $x_{i}^t=0$ is the resting state, $x_{i}^t=1$ is the excited state, and there may be additional refractory states $x_{i}^t=2,3,...,m_i$. 
\item  If element $i$ is in the resting state at time $t$, $x_{i}^{t}=0$, it can be excited in the next time step, $x_{i}^{t+1}=1$, by another excited element $j$ with delay $\tau_{ij}$ (i.e., if $x_{j}^{t-\tau_{ij}}=1$) with probability $A_{ij}$, or independently by an external stimulus with probability $\eta$.
\item The elements that are excited or in a refractory state, $x_{i}^{t} \geq 1$, will deterministically make a transition to the next refractory state if one is available, or return to the resting state otherwise (i.e., $x_{i}^{t+1} = x_{i}^{t} + 1$ if $1 \leq x_{i}^{t} < m_i$, and $x_{i}^{t+1} = 0$ if $x_{i}^{t} = m_i$).
\item The coupling network, encoded by the matrix with entries $A_{ij}$, is allowed to have complex topology. 
\end{itemize}

\subsection{Model dynamics}

By considering a large ensemble of realizations of the above stochastic process on the same network, we can define the probability that node $i$ is at state $x_i^t$ at time $t$ as $p_i^t(x)$. The probabilities $p_i^{t}$ evolve in one time step by
\begin{align}
p_i^{t+1}(1)& = p_i^t(0)r_i^t, \label{mass}\\
p_i^{t+1}(2) &= p_i^t(1),\\
&\cdots\\
p_i^{t+1}(m_i) &= p_i^t(m_i-1),
\end{align}
and we also have the normalization condition
\begin{align}
p_i^{t}(0) &= 1- \sum_{j=1}^{m_i}p_i^t(j),
\end{align}
where $r_i^t$ in Eq.~(\ref{mass}) is the rate of transitions from the ready to the excited state, given by
\begin{align}
r_i^t =  E\left[ \eta + (1-\eta)\left\{ 1 - \prod_{j} (1-A_{ij}I_j^{t-\tau_{ij}})       \right\}  \right],
\end{align}
where $I_j^t$ is one if node $j$ is excited at time $t$ and zero otherwise, and $E[\cdot]$ denotes an ensemble average. Assuming that the neighbors of node $i$ being excited are independent events, we obtain, letting $p_i^t(1) \equiv p_i^t$,
\begin{align}\label{needindependence}
r_i^t =  \eta + (1-\eta)\left\{ 1 - \prod_{j} (1-A_{ij}p^{t-\tau_{ij}}_j)       \right\}.
\end{align}
We note that the assumption of independence is reasonable if there are few short loops in the network, and has been successfully used in similar situations \cite{locallyTreeLike,genes}. However, this assumption is violated if the number of bidirectional links is significant, and therefore we will restrict our attention to purely directed networks.
Inserting the expression above in Eq.~(\ref{mass}) and eliminating $p_i^t(j)$ in terms of $p_i^t$  for $j = 2,\dots,m_i$, we obtain the governing equation for the dynamics of $p_i^t$
\begin{widetext}
\begin{equation}\label{nasty}
p_{i}^{t+1} =  \left(1 - \sum_{k=1}^{m_i} p_{i}^{t+1-k}\right)\left( \eta + (1-\eta) \left [ 1- \prod_{j}^{N} (1 -  A_{ij}p_{j}^{t-\tau_{ij}}) \right ] \right).
\end{equation}
\end{widetext}
In the following, we will analyze the response of the network by studying solutions of this equation as a function of $\eta$.

\section{Analysis}\label{analysis}

In this Section we study the solutions of Eq. (9) and the associated network response. In Sec. IV A, we develop a nonperturbative approximation to the steady state response of the network. In Sec. IV B we analyze limiting cases of the steady-state response that give us additional qualitative insight. In Sec. IV C, we study the effect of a distribution in the transmission time delays on the time scale of relaxation to the steady state solutions. We then discuss in Sec. IV D how our results relate to previous work.

\subsection{Steady-state response}\label{nonperturbative}

First, we will study steady-state solutions to Eq.~(\ref{nasty}). To find those, we set $p_i^t =p_i$ in  Eq.~(\ref{nasty}), which becomes
\begin{equation}\label{expstate}
p_{i} =  \left(1 - m_i p_{i}\right)\left( \eta + (1-\eta) \left [ 1- \prod_{j}^{N} (1 -  A_{ij}p_{j}) \right ] \right).
\end{equation}
Proceeding as in Ref.~\cite{Larremore}, by assuming $A_{ij} p_j$ is small, we replace $\prod_{j}(1-A_{ij}p_j)$ by $\exp(-\sum_{j}{A_{ij}p_j)}$ to get
\begin{equation}\label{sstate}
p_{i} =  \left(1 - m_ip_{i}\right)\left( \eta + (1-\eta) \left [ 1- e^{-\sum_j A_{ij}p_{j}}) \right ] \right).
\end{equation}
The assumption that $A_{ij}p_j$ is small is motivated as follows. If the weights $A_{ij}$ are not very different from each other and each node has many incoming connections (such as in neural networks, where the number of synapses per neuron is estimated\cite{neuralConnectivityBook} to be of the order of $10,000$), then near the onset of self-sustained activity one should have $\sum_j A_{ij} \sim 1$ (the mean-field prediction of Ref.~\cite{kinouchiCopelli}, which we refine here, states that the node average of $\sum_j A_{ij}$ is one at criticality), implying $A_{ij}$ is small. The quantity $A_{ij}p_j$ is even smaller, especially for low levels of activity where $p_j$ is small.

To proceed further we find convenient to define an alternative network response $\hat F$ as
\begin{align}\label{efe}
\hat F = \langle \hat f \rangle_t ,
\end{align}
where 
\begin{align}\hat f^{t} = \frac{\sum_{i,j} A_{ij}I_{j}^{t}}{\sum_{i,j} A_{ij}},
\end{align}
and $I_{j}^{t} = 1$ if node $j$ is excited at time $t$ and $0$ otherwise. The variable $\hat f^{t}$ can be interpreted as proportional to  the number of excited nodes weighted by their out-degree $d^{out}_{j} = \sum_{i} A_{ij}$. In terms of the probabilities $p_{i}$, $\hat F$ is 
\begin{equation}
\hat F = \frac{\sum_{i,j} A_{ij}p_j}{\sum_{i,j} A_{ij}},
\end{equation}
and can be interpreted as the fraction of links that successfully transmit an excitation. This is analogous to the interpretation of $F$ in Eq.~(\ref{response}) as the fraction of excited nodes. In principle, the definitions of $\hat{f}$ and $\hat{F}$ preclude their use in comparing directly against  commonly used measures of activity since knowledge of the matrix $A$ is required to estimate them. However, we note that in all the numerical experiments discussed below, $\hat{F}$ and $F$ were found to be nearly identical. To develop a nonperturbative approximation to $\hat F$, we solve Eq.~(\ref{sstate}) for $p_i$ in terms of $\sum_{j=1}^N A_{ij} p_j$. Multiplying the resulting expression by $A_{ki}$ and summing over $i$, we obtain
\begin{equation}
\sum_{i=1}^N A_{ki} p_i = \sum_{i=1}^N A_{ki}\frac{1 - (1-\eta) e^{-\sum_{j=1}^NA_{ij}p_j}}{1 + m_i - m_i(1-\eta) e^{-\sum_{j=1}^NA_{ij}p_j}}.
\end{equation}
Now, we use the fact that the largest eigenvalue of $A$, $\lambda$, is typically much larger than the second eigenvalue\cite{eigenvalueApproximation, chauhanOtt}, and thus $Ap \approx s u$, where $s$ is a scalar to be determined, and $u$ is the right eigenvector of $A$ corresponding to $\lambda$. The validity of this approximation will be discussed in Section \ref{validity}. With this substitution, the previous equation reduces to 
\begin{equation}\label{nonpek}
s u_k = \sum_{i=1}^N A_{ki}\frac{1 - (1-\eta) e^{-s u_i}}{1 +m_i - m_i(1-\eta) e^{-s u_i}}.
\end{equation}
Noting that 
$$
\hat F = \frac{\sum_{i,j} A_{ij}p_j}{ \sum_{i,j} A_{ij}}\approx \frac{\sum_i s u_i }{\sum_j d_j^{out}} =  s \frac{\langle u \rangle}{\langle d \rangle},
$$  
where $\langle x \rangle \equiv \sum_{i=1}^N x_i/N$, we substitute $s \approx \hat{F} \langle d \rangle / \langle u \rangle$ into Eq. (\ref{nonpek}) yielding
$$
\frac{\hat{F} u_k \langle d \rangle}{\langle u \rangle}  = \sum_{i=1}^N A_{ki}\frac{1 - (1-\eta) e^{- \hat{F}  u_i \langle d \rangle / \langle u \rangle}}{1 +m_i - m_i(1-\eta) e^{-\hat{F} u_i \langle d \rangle / \langle u \rangle }}.
$$
which may now be summed over $k$, simplified, and solved for $\hat{F}$, yielding the scalar equation
\begin{equation}\label{nonpe}
\hat F = \left\langle \frac{d^{out}}{\langle d\rangle} \frac{1 - (1-\eta) e^{-\hat F u\langle d \rangle/\langle u\rangle}}{1 + m - m(1-\eta) e^{-\hat F u\langle d \rangle/\langle u\rangle}}\right \rangle.
\end{equation}
We note that in the notation above, the outer average $\langle \cdot \rangle$ corresponds to a sum over the index $i$ in Eq.~(\ref{nonpek}).
Given the adjacency matrix $A$, Eq. (\ref{nonpe}) can be solved numerically to obtain the response $\hat F$ as a function of $\eta$. We call Eq.~(\ref{nonpe}) the ``nonperturbative approximation'' since its derivation does not rely on a perturbative truncation of the product term of Eq.~(\ref{expstate}), and we will numerically test its validity in Sec.~\ref{numerical}, where we will find that Eq.~(\ref{nonpe}) can be a good approximation for all values of $\eta$. In order to gain theoretical insight into how some features of the network topology and the distribution of the number of refractory states affect the response, we will use Eq.~(\ref{nonpe}) to obtain analytical expressions for the response in various limits.

\subsection{Perturbative approximations}\label{perturbative}

While the nonperturbative approximation developed in the last section provides information for all ranges of stimulus, it is useful to consider perturbative approximations, for example, to determine what is the transition point from quiescent to active behavior.
We will obtain an approximation to $\hat F$ which is valid for small $\eta$ and $\hat F$. To do this, we expand the right hand side of Eq.~(\ref{nonpek}) to second order in $s$ and first order in $\eta$ (as we will see, expanding to second order in $s$ is necessary to treat the $\eta = 0$ case) obtaining
\begin{align}
&s u_k = 
\\&\sum_{i=1}^N A_{ki}\left[ \eta (1 - su_i(1+2m_i)) -s^2 u_i^2\left(\frac{1}{2} +m_i\right) +s u_i   \right].\nonumber
\end{align}
Multiplying by the left eigenvector entry $v_k$ and summing over $k$ we obtain, using $\sum_k A_{ki}v_k =\lambda v_i$ and rearranging,
\begin{align}
\lambda s^{2} \left\langle v u^2\left(\frac{1}{2} +m\right)\right\rangle &=  \eta\lambda  \langle v \rangle - s \eta\lambda \langle vu(1+2m)\rangle \\ 
& + (\lambda-1) s \langle vu\rangle\nonumber.
\end{align}
In terms of $\hat F$, this equation becomes
\begin{align}\label{fEq}
&\hat F^2 \langle d \rangle^2\left\langle v u^2\left(\frac{1}{2} +m\right)\right\rangle \lambda =   \eta\lambda  \langle v \rangle \langle u \rangle^2 \\ &- \hat F \langle d \rangle\eta\lambda \langle vu(1+2m)\rangle\langle u \rangle
+ (\lambda-1) \hat F \langle d \rangle\langle vu\rangle \langle u \rangle\nonumber.
\end{align}

To find the transition from no activity, $\hat F = 0$, to self-sustained activity, $\hat F > 0$, for vanishing stimulus, we let $\eta \to 0^+$ in the previous equation to find
\begin{align}\label{fbarzero}
 \hat F_{\eta = 0}(\lambda) = \left\{ 
  \begin{array}{l l}
    0 & \lambda < 1,\\
   \frac{(\lambda-1)\langle vu \rangle \langle u \rangle}{\lambda \langle d \rangle \left\langle v u^2\left(m + \frac{1}{2}\right)\right\rangle} & \lambda \geq 1,
  \end{array} \right.
  \end{align}
where the solution $\hat F  = 0$ was chosen for $\lambda < 1$ to satisfy $\hat F \geq 0$.
This equation shows that the transition from a quiescent network to one with self-sustained activity has, if the response $\hat F$ is interpreted as an order parameter, the signatures of a second order (continuous) phase transition. In addition, while the eigenvalue $\lambda$ determines when this transition occurs (at $\lambda = 1$), its associated eigenvectors $u$ and $v$ determine the significance of the observed response past the transition. If $\langle v u^{2}  \rangle \gg \langle uv \rangle \langle u \rangle$, for example, the response might be initially too small to be of  importance. One aspect that was not considered in Ref.~\cite{Larremore} is how the distribution of refractory periods affects the response. If the refractory periods $m_i$ are strongly positively correlated with the product $v_i u_i^2$, they can significantly increase the term $\langle m v u^{2} \rangle$ in the denominator, decreasing the response. This can be intuitively understood by noting that this amounts to preferentially increasing the refractory period of the nodes that are more likely to be active (as measured by the approximation $p_i \propto u_i$  valid close to the critical regime), thus removing them from the available nodes for longer times.

The response $\hat F$ for small stimulus and response in Eq.~(\ref{fbarzero}) agrees with the perturbative expression derived for $F$ directly from Eq.~(\ref{nasty}) in Ref.~\cite{Larremore} if $m_i = 1$ and $\lambda \to 1$, and confirms the findings in Ref.~\cite{Larremore} that the critical point is determined by $\lambda = 1$. Henceforth, we will refer to networks with $\lambda < 1$ as \textit{quiescent}, to networks with $\lambda > 1$ as \textit{active}, and to networks with $\lambda=1$ as \textit{critical}.

The behavior of the system for high stimulation is also of interest. When $\eta=1$, node $i$ cycles deterministically through its $m_i+1$ available states, and so $p_i = (1+m_i)^{-1}$. The question is how this behavior changes as $\eta$ decreases from $1$. This information can be extracted directly from Eq.~(\ref{sstate}) by linearizing around the solution  $\eta = 1$, $\bar p_i = (1+m_i)^{-1}$. Setting $\eta = 1 - \delta \eta$, $p_i = \bar p_i -\delta p_i$ with $\delta \eta \ll 1$, $\delta p _i \ll \bar{p_i}$, we obtain
\begin{align}\label{deltapi}
\delta p_i \approx   \bar p_i^2 e^{-\sum_j A_{ij}\bar p_j} \delta \eta.
\end{align}
Thus, the response of the nodes to a decreased stimulus depends on a combination of their refractory period (which determines $\bar p_i$) and decays exponentially with the number of expected excitations from its neighbors. In terms of the aggregate response $\hat{F}$, Eq.~(\ref{deltapi}) becomes, after multiplting by $A_{ki}$, summing over $k$ and $i$, and normalizing,
\begin{align}\label{dFdeta}
\frac{d \hat{F}} {d \eta} = \frac{\langle d^{out} \bar{p}^{2} e^{-A \bar{p}} \rangle}{\langle d \rangle}
\end{align}

\subsection{Dynamics near the critical regime}\label{dynamics}

As in Ref.~\cite{Larremore}, we will study the transition from no activity to self-sustained activity in the limit of vanishing stimulus by linearizing Eq.~(\ref{nasty}) around $p_i^t = 0$ for $\eta = 0$. Assuming $p_i^t$ is small, we obtain to first order
\begin{equation}\label{delayeq}
p_{i}^{t+1} =\sum_{j=1}^{N} A_{ij} p_{j}^{t-\tau_{ij}}.
\end{equation}
Assuming exponential growth, $p_i^t = \alpha^t w_i$, we obtain
\begin{equation}\label{alphas}
\alpha w_{i} =\sum_{j=1}^{N}A_{ij} \alpha^{-\tau_{ij}}w_{j}.
\end{equation} 
The critical regime, determined as the boundary between no activity and self-sustained activity as $\eta \to 0$, i.e, between the  
solution $p_i^t = 0$ being stable and unstable, can be found by setting $\alpha = 1$, obtaining
\begin{equation}
w_{i} =\sum_{j}^{N} A_{ij} w_{j}.
\end{equation}
This implies that the onset of criticality occurs when $\lambda = 1$ and in this case $w = u$. This conclusion agrees with the results in Ref.~\cite{Larremore} and those in the previous section. Although the critical regime is not affected by the presence of delays or refractory states, the rate of growth (decay) $\alpha$ of perturbations  for the active (quiescent) regime depends on the distribution of delays. To illustrate this, we consider the case when the network deviates slightly from the critical state, so that the largest eigenvalue of $A$ is $\lambda = 1 +\delta\lambda$ and has right eigenvector $u$, $A u = (1+\delta \lambda) u$. Expecting the solution $w$ to Eq.~(\ref{alphas}) to be close to $u$, we set  $w_i =  u_i + \delta u_i$  and $\alpha = 1 + \mu$, where the rate of growth $\mu$ is assumed to be small. Inserting these in Eq.~(\ref{alphas}), we get to first order
\begin{align}
\mu  u + \delta u \approx  u\delta\lambda   + A \delta u - \mu \hat A  u,
\end{align}
where the entries of the matrix $\hat A$ are given by $\hat A_{ij} = A_{ij}\tau_{ij}$. To eliminate the term $\delta u$, we left-multiply by the left eigenvector of $A$, $v$, satisfying $v^T A = (1+\delta\lambda)  v^T $. Canceling small terms, we get
\begin{align}\label{newgrowthrate}
\mu \approx \frac{\delta \lambda}{1 + \frac{ v^T\hat A  u}{ v^T u}}.
\end{align}
If the delay is constant, $\tau_{ij} = \tau$, we obtain
\begin{align}
\mu \approx \frac{\delta \lambda}{1 + \tau},
\end{align}
and in this particular case a more general result can be obtained from Eq.~(\ref{alphas}), which implies $\alpha = \lambda^{1/(1+\tau)}$.
\subsection{Relation to previous results}

Here we will briefly discuss how our results for the critical regime agree with previous work in particular cases. Correlations between degrees at the ends of a randomly chosen edge (assortative mixing by degree \cite{newmanAssortativity}) can be measured by the correlation coefficient 
\begin{align}\label{rhoDef}
\rho = \langle d_{i}^{in} d_{j}^{out} \rangle_{e}/\langle d^{in} d^{out} \rangle
\end{align}
 defined in Ref.~\cite{eigenvalueApproximation}, with $\langle \cdot \rangle_{e}$ denoting an average over edges.  The correlation coefficient $\rho$ is greater than $1$ if the correlation between the in-degree and out-degree of nodes at the end of a randomly chosen edge is positive, less than one if the correlation is negative, and one if there is no correlation. For a large class of networks, the largest eigenvalue may be approximated\cite{eigenvalueApproximation} by $\lambda \approx \rho \langle d^{in} d^{out} \rangle / \langle d \rangle$. In the absence of correlations, when $\rho = 1$, the largest eigenvalue can be approximated by $\lambda \approx \langle d^{in} d^{out} \rangle/\langle d \rangle$. If there are no correlations between $d^{in}$ and $d^{out}$ at a node (\textit{node degree correlations}) or if the degree distribution is sufficiently homogeneous, then $\langle d^{in} d^{out} \rangle \approx \langle d \rangle^{2}$ and the approximation reduces to $\lambda \approx \langle d \rangle$. This is the situation that was considered in Ref. \cite{kinouchiCopelli}, and thus they found that the critical regime was determined by $\langle d \rangle = 1$. In the case of Refs. \cite{copelliCampos,wu}, with more heterogeneous degree distributions, $\lambda \approx \langle d^{in} d^{out} \rangle/\langle d \rangle$ applied, which accounts for their observation that the critical regime did not occur at $\langle d \rangle = 1$.

The situation encountered here is analogous to what occurs in the analysis of the transition to chaos in Boolean networks \cite{genes} and in the transition to synchronization in networks of coupled oscillators \cite{kuramotoNetworks}, where it is found that, instead of the mean degree, the largest eigenvalue is what determines the transition between different collective dynamical regimes.

Other previous studies in random networks have also investigated spectral properties of $A$ to gain insight on the stability of dynamics in neural networks\cite{gray} and have shown how $\lambda$ could be changed by modifying the distribution of synapse strengths\cite{rajan}. In addition, it has been shown recently that the largest eigenvalues in the spectrum of the connectivity matrix may affect learning efficiency in recurrent chaotic neural networks \cite{sussillo}.

\section{Dynamic Range}\label{dynamicrange}

\begin{figure}[h]
\centering \epsfig{file =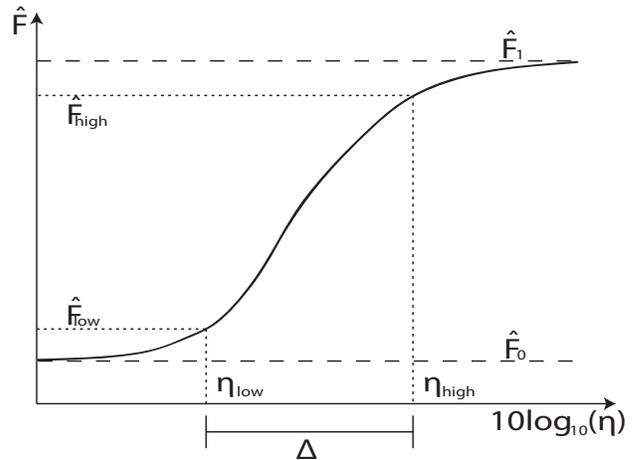, clip =
,width=1.0\linewidth }
\caption{Schematic illustration of the definition of dynamic range in the active network case. The baseline and saturation values are $\hat F_0$ and $\hat F_1$, respectively. Two threshold values, denoted by $\hat F_{low}$ and $\hat F_{high}$, respectively, are used to determine the range of values of $\eta$ defined as the {\it dynamic range} $\Delta$. 
} \label{maxDRfig}
\end{figure}

We have studied the response of the network to stimuli of varying strengths. In particular, we studied in detail the response close to the critical regime. As has been previously noted \cite{kinouchiCopelli}, this regime corresponds to the point where the {\it dynamic range} $\Delta$ is maximized. In our context, the dynamic range can be defined as the range of stimulus $\eta$ that results in significant changes in response $\hat F$. Typically the dynamic range is given in decibels and measured using arbitrary thresholds just above the baseline ($\lim_{\eta \to 0} \hat F \equiv \hat F_0$) and below the saturation ($\lim_{\eta\to1} \hat F \equiv \hat F_1$) values, respectively, as illustrated in Fig.~\ref{maxDRfig} for the active network case $\hat F_0 > 0$. More precisely, the value of stimulus $\eta_{low}$ ($\eta_{high}$) corresponding to a low (high) threshold of activity $\hat F_{low}$ ($\hat F_{high}$) are found and the dynamic range is calculated as 
\begin{align}\label{drange}
\Delta = 10 \log_{10}(\eta_{high}/\eta_{low}).
\end{align}

Using our approximations to the response $\hat F$ as a function of stimulus $\eta$, we can study the effect of network topology on the dynamic range. The first approximation is based on the analysis of Sec.~\ref{nonperturbative}. Using Eq.~(\ref{nonpe}), the values of $\eta$ correponding to a given stimulus threshold can be found numerically and the dynamic range calculated.

Another approximation that gives theoretical insight into the effects of network topology and the distribution of refractory states on the dynamic range can be developed as in Ref.~\cite{Larremore}, by using the perturbative approximations developed in Sec.~\ref{perturbative}. In order to satisfy the restrictions under which those approximations were developed, we will use $\hat F_{high} = \hat F_1$ and $\hat F_{low} = \hat F_0\ll 1$. Taking the upper threshold to be $\hat F_{high} = \hat F_1$ is reasonable if the response decreases quickly from $\hat F_1$, so that the effect of the network on the dynamic range is dependent mostly on its effect on $\hat F_{low}$. Whether or not this is the case can be established numerically or theoretically from Eq.~(\ref{deltapi}), and we find it is so in our numerical examples when $m_{i}$ are not large (see Fig.~\ref{topSlope}). Taking $\eta_{high} = 1$ and $\eta_{low} = \eta^*$ we have
\begin{align}\label{delta}
\Delta = -10 \log_{10}(\eta^*).
\end{align}
The stimulus level $\eta$ can be found in terms of $\hat F $ by solving Eq.~(\ref{fEq}) and keeping the leading order terms in $\hat F$, obtaining
\begin{equation}\label{FzeroEq2}
\eta = \frac{ \hat  F^2 \langle d \rangle^2\left\langle v u^2\left(\frac{1}{2} +m\right)\right\rangle- \hat F \langle d \rangle (\lambda - 1) \langle u  \rangle\langle u v \rangle}{\lambda  \langle v \rangle \langle u \rangle^2}.
\end{equation} 
This equation shows that as $\eta \to 0$ the response scales as $\hat F \sim \eta$ for the quiescent curves ($\lambda < 1$), and as $\hat F \sim \eta^{1/2}$ for the critical curve ($\lambda = 1$). We highlight that these scaling exponents for both the quiescent and critical regimes are precisely those derived in Ref.~\cite{kinouchiCopelli} for random networks, attesting to their robustness to the generalization of the criticality criterion to $\lambda = 1$, the inclusion of time delays, and heterogeneous refractory periods . This is particularly important since these exponents could be measured experimentally~\cite{kinouchiCopelli}.
Using this approximation for $\eta^*$ in (\ref{delta}), we obtain an analytical expression for the dynamic range valid when the lower threshold $F^*$ is small. 
Of particular theoretical interest is the maximum achievable dynamic range $\Delta_{max}$ for a given topology. It can be found by setting $\lambda = 1$ in Eq.~(\ref{FzeroEq2}) and inserting the result in Eq.~(\ref{delta}), obtaining
\begin{align}\label{DR}
\Delta_{\mbox{max}} = \Delta_0- 10\log_{10} \left(\frac{ \langle d \rangle^2\left\langle v u^2\left(\frac{1}{2} +m\right)\right\rangle}{ \langle v \rangle \langle u  \rangle^2 }\right),
\end{align}
where $\Delta_0 = -20\log_{10}(F^*) > 0$ depends on the threshold $F^*$ but is independent of the network topology or the distribution of refractory states. The second term of Eq. (\ref{DR}) suggests that a positive correlation between refractoriness $m$ and eigenvector entries $u$ and $v$ will decrease dynamic range, whereas a negative correlation will increase dynamic range. This prediction may be investigated in more depth in future publications. The second term of Eq. (\ref{DR}) also suggests that an overall increase in the number of refractory states will lead to an overall decrease in dynamic range. This is in contrast with the result of Ref. \cite{copelliRefractory}, which found that there exists a $m>0$ which maximizes dynamic range in two-dimensional arrays of neurons. We note that the assumption of independence used in deriving Eq.~(\ref{needindependence}) is not valid for a two dimensional array.

\section{Numerical experiments}\label{numerical}

\begin{figure*}[t]
\centering
\epsfig{file =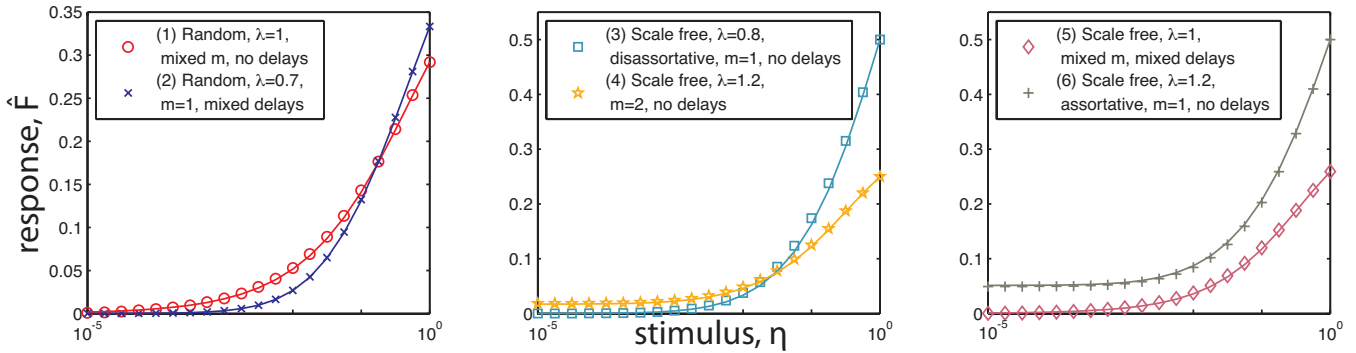, clip=, width=1\linewidth}
\caption{Semi-log plots of data from five simulations (symbols) testing a variety of situations in order to show the robustness of Eq.~(\ref{nonpe}) (lines) to various sets of conditions: 
(1) Random network; $\lambda=1$; mixed refractory states, $m_{i} \in \{1,2,3\}$; no delays; 
(2) Random network; $\lambda=0.7$; no refractory states, $m_{i}=m=1$; mixed delays, $\tau_{ij} \in \{0,1,2,3\}$;
(3) Scale free network; $\lambda=0.8$; disassortative rewiring; no refractory states, $m_i=m=1$; no delays;
(4) Scale free network; $\lambda=1.2$; uniform refractory states, $m_{i}=m=2$; no delays;
(5) Scale free network; $\lambda=1.0$; mixed refractory states, $m_{i} \in \{1,2,3,4\}$; mixed delays, $\tau_{ij} \in \{0,1,2\}$;
(6) Scale free network; $\lambda=1.2$; uniform refractory states, $m_{i}=m=1$; no delays. The plots show excellent agreement between prediction and simulation at many points in parameter space.}\label{semilogzoo}
\end{figure*}

We tested our theoretical results from section \ref{analysis} by comparing their predictions to direct simulation of our generalized Kinouchi-Copelli model described in section \ref{longtitle}. Simulation parameters were chosen specifically to test the validity of Eqs.~(\ref{nonpe}), (\ref{dFdeta}), (\ref{newgrowthrate}) and (\ref{DR}). All simulations, except where indicated,  were run with $N=10^{4}$ nodes for $T=10^{5}$ timesteps, over a range of $\eta$ from $10^{-5}$ to $1$.

\subsection{Construction of networks}

We created networks in three steps: first, we created binary directed networks, $A_{ij} \in \{0,1 \}$, with particular degree distributions as described below, forbidding bidirectional links and self-connections; second, we assigned a weight to each link, drawn from a uniform distribution between 0 and 1; third, we calculated $\lambda$ for the resulting network, and multiplied $A$ by a constant to rescale $\lambda$ to the targeted eigenvalue\cite{endnote1}. The two classes of topology considered for simulations were directed Erd\H{o}s-R\'{e}nyi random networks and directed scale-free networks with power law degree distributions, where we set the power law exponent to $\gamma=2.5$, and enforced a minimum degree of 10 and a maximum degree of 1000. Erd\H{o}s-R\'{e}nyi networks\cite{ERGraph} were constructed by linking any pair of nodes with probability $p = 15/N$, and scale-free networks were constructed by first generating in-degree and out-degree sequences drawn from the power law distribution described above, assigning those target degrees to $N$ nodes, and then connecting them using the configuration model \cite{configurationModel}. In some cases an additional fourth step was used to change the assortativity coefficient $\rho$, defined in Eq.~(\ref{rhoDef}), of a critical (i.e., with $\lambda = 1$) scale-free network, making this network more assortative (disassortative) by choosing two links at random, and swapping their destination connections only if the resulting swap would increase (decrease) $\rho$. This swapping allows for the degree of assortativity (and thereby, $\lambda$) to be modified while preserving the network's degree distribution \cite{eigenvalueApproximation,newmanAssortativity}.

\subsection{Results of numerical experiments}

We first demonstrate the ability of the non-perturbative approximation to predict aggregate network behavior in a variety of conditions. Fig. \ref{semilogzoo} shows a multitude of simulations (symbols) with the predicted behavior of Eq.~(\ref{nonpe}) overlaid (lines). The cases considered in Fig.~\ref{semilogzoo} include different combinations of topology, assortativity, largest eigenvalue $\lambda$, delays, and number of refractory states. The number of refractory states $m_i$ was chosen either constant, $m_i = m$, or randomly chosen with equal probability among $\{1,2,\dots,m_{\text{max}}\}$. Similarly, the delays $\tau_{ij}$ were either constant, $\tau_{ij} = \tau$, or uniformly chosen with uniform probability in $(0,\tau_{\text{max}})$. The predictions capture the behavior of the simulations, with particularly good agreement for networks with neutral assortativity, $\rho=1$. In the assortative and disassortative cases shown [cases (3) and (6) in Fig.~\ref{semilogzoo}], low and high stimulus simulations are well captured by the prediction, while a small deviation can be observed for intermediate values of $\eta$ [e.g., in case (6) in the right panel of Fig.~\ref{semilogzoo}, the crosses have a small systematic error around $\eta = 10^{-2}$]. In Sec.~\ref{validity}, we will discuss why Eq.~(\ref{nonpe}), which assumes $Ap \approx su$, works so well. In particular, we will discuss why this approximation is expected to work well for small and large $\eta$.

\begin{figure}[t]
\centering
\subfigure{
\epsfig{file =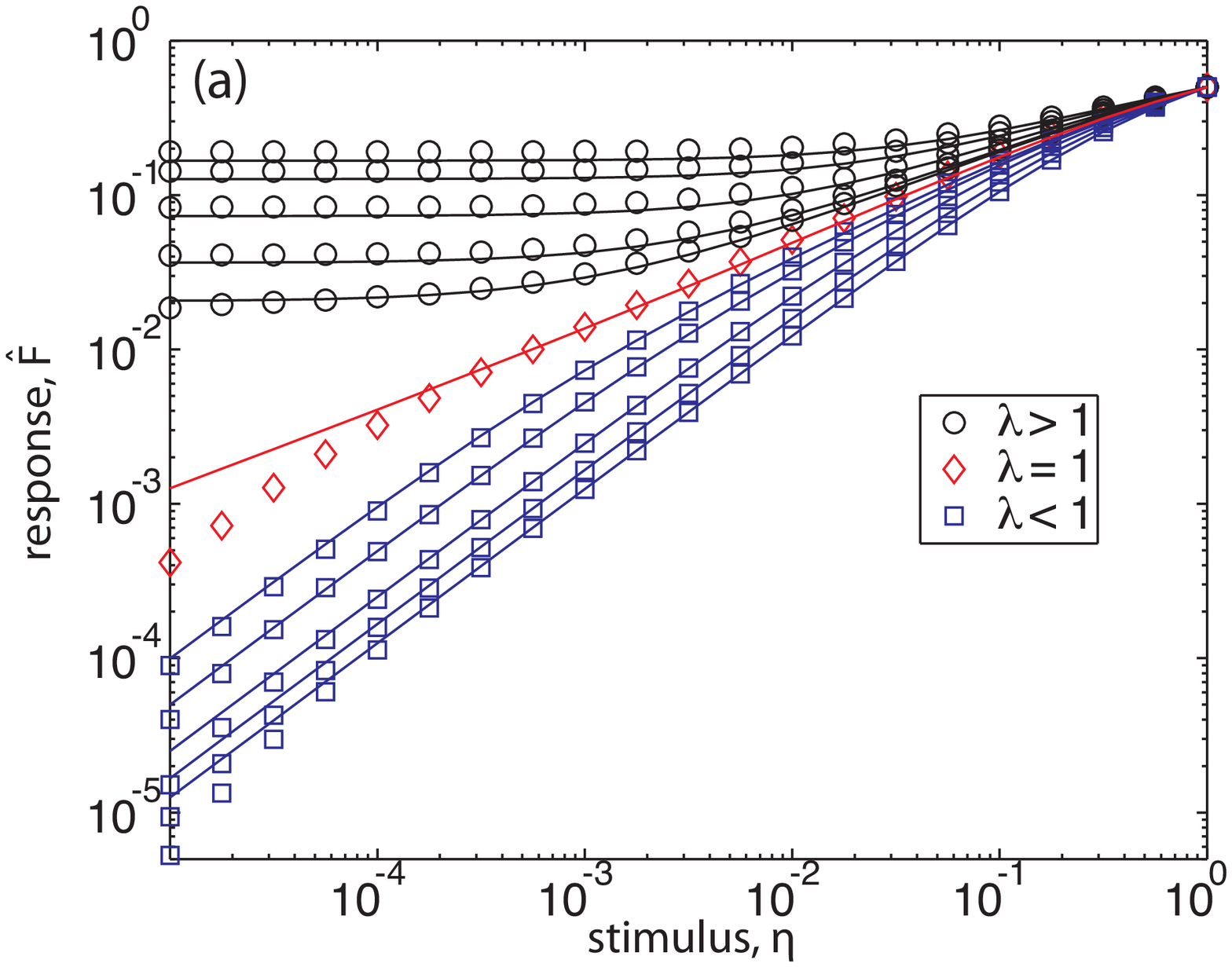, clip =,width=1.0\linewidth }
}
\subfigure{
\epsfig{file =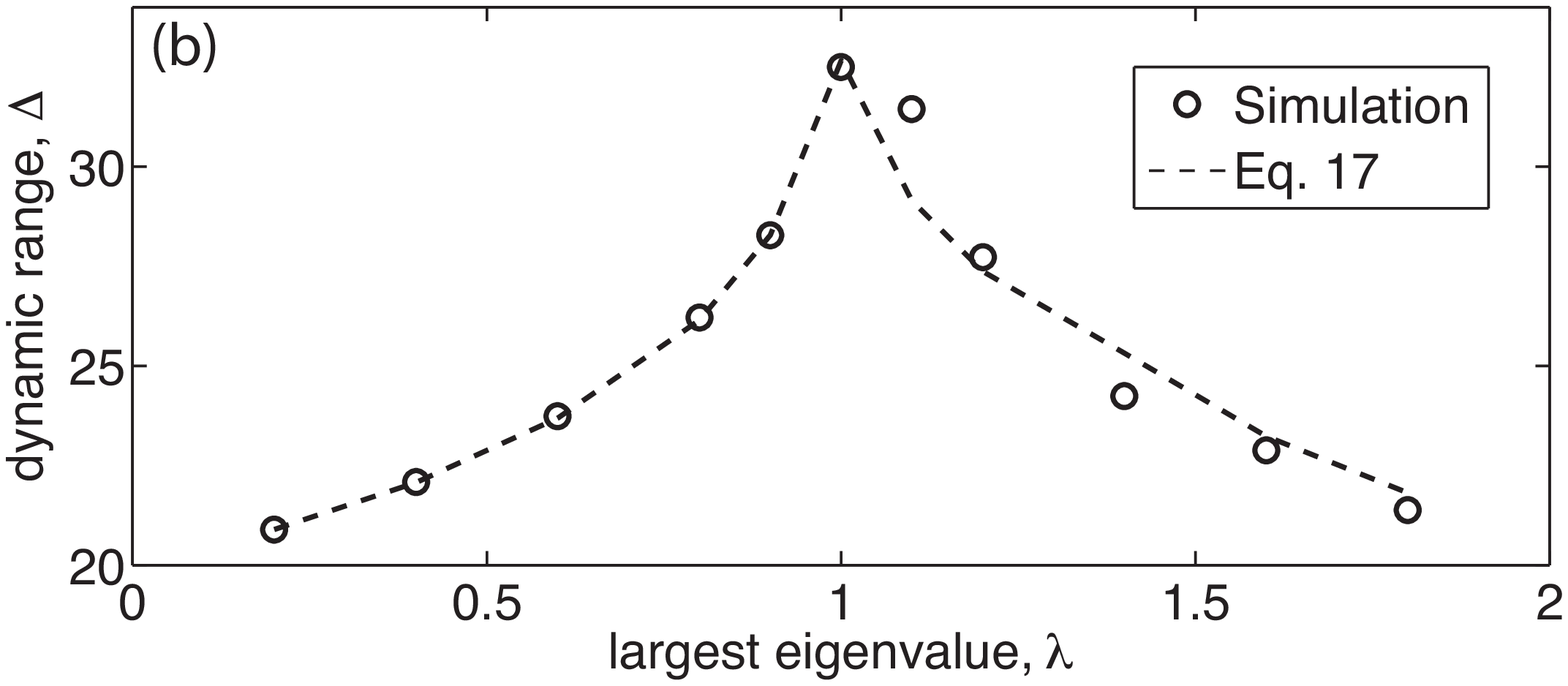, clip=,width=1.0\linewidth}
}
\caption{Simulation data for scale-free networks of $10^{4}$ nodes (symbols) and numerical solution of Eq. \ref{nonpe} (lines). (a) Stimulus vs response predictions agree well in the regime where $Ap \approx su$, as discussed in section \ref{validity}. Eigenvalues range from 0.2 to 0.9 (blue squares), exactly 1.0 (red diamonds), and from 1.1 to 1.8 (black circles). (b) Dynamic range predictions capture maximization at $\lambda=1$ as well as the non-critical behavior.} \label{stimulusResponseCurves}
\end{figure}

As reported previously\cite{Larremore}, we find in our simulations that networks with $\lambda=1$ show critical dynamics and exhibit maximum dynamic range. This applies to random networks, scale free networks, and scale free networks with modified assortativity. Networks with $\lambda < 1$ exhibit no self-sustained activity in the absence of stimulus, whereas networks with $\lambda > 1$ exhibit self-sustained activity. Furthermore, in all numerical experiments, with distributed refractory states, and various time delays, the criticality of networks at $\lambda=1$ was preserved as predicted above. Typical results in Fig.~\ref{stimulusResponseCurves} (a) show the response $\hat{F}$ as a function of stimulus $\eta$ for scale free networks with $\gamma=2.5$, refractory states $m_i = m = 1$, and no time delays, with $\lambda$ ranging from 0.2 to 1.8. Each symbol in the figure is generated by a single simulation on a single network realization. Lines show $\hat F$ obtained from numerical solution of Eq.~(\ref{nonpe}). We note that the simulations with $\lambda = 1$ show a deviation from the theoretically predicted critical curve for values of $\eta$ less than $10^{-4}$. We believe this is due to the fact that for such low values of $\eta$, a much longer time average than the one we are doing would be required. For example, with $\eta=10^{-5}$ we expect that, using $10^5$ time steps, a given node will not be excited externally with probability $e^{-1}\approx 0.37$. This might be especially important in the critical regime, where activity is mostly determined by sporadic avalanches propagated by hubs.

\begin{figure}[h]
\epsfig{file =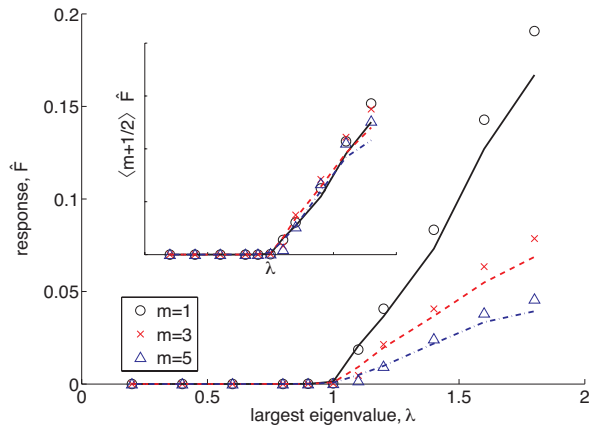,width=1.0\linewidth }
\caption{Phase transitions of $F_{\eta \to 0}$ for different refractory states, $m$ for simulations (symbols) and Eq. (\ref{nonpe}) (lines). Inset: Eq. (\ref{fbarzero}) predicts that phase transitions should scale by $\langle m + 1/2 \rangle^{-1}$, confirmed by rescaling data from the larger plot accordingly.}\label{phaseTransitions}
\end{figure}

Figure~\ref{stimulusResponseCurves} (b) shows the dynamic range $\Delta$ calculated using $F^* = 10^{-2}$ directly from the simulation (circles) and using Eq.~(\ref{nonpe}) (dashed line). As demonstrated in Ref.\cite{Larremore}, the dynamic range is maximized when $\lambda = 1$. We note that in Ref.\cite{Larremore} the dynamic range was estimated using a perturbative approximation, and as a consequence our prediction had a systematic error in the $\lambda > 1$ regime [cf. Fig.~1 (b) in Ref.\cite{Larremore}]. The nonperturbative approximation Eq.~(\ref{nonpe}) results in a much better prediction.

Figure~\ref{phaseTransitions} shows the transition that occurs at $\lambda = 1$ when $\eta \to 0$ for experiments with a varying number of refractory states, $m=1$, $3$, and $5$. Symbols indicate the results of direct simulation using $\eta = 10^{-5}$, and the lines correspond to Eq.~(\ref{nonpe}), which describes well the result of the simulations. We found that for this particular network, the perturbative approximation (\ref{fbarzero}) only gives correct results very close to the transition at $\lambda = 1$, and its quantitative predictions degrade quickly as $\hat F$ grows. [A similar situation can be observed in Fig.~2(b) of Ref.\cite{Larremore}.] However, we found that the perturbative approximation is still  useful to predict the effect of the refractory states. Eq.~(\ref{fbarzero}) predicts that the response should scale as $\langle m+1/2\rangle^{-1}$. The inset shows how, after multiplication by $\langle m+1/2\rangle$, the response curves collapse into a single curve. Figure~\ref{phaseTransitions} also depicts a linear relationship, $\hat{F} \sim (\lambda - 1)$ for $\lambda > 1$. Making a connection with the theory of nonequilibrium phase transitions in which $\hat{F} \sim (\lambda - \lambda_c)^{\beta}$, we derive $\lambda_c = 1$ and the critical exponent $\beta = 1$.
 
\begin{figure}[h]
\centering
\epsfig{file =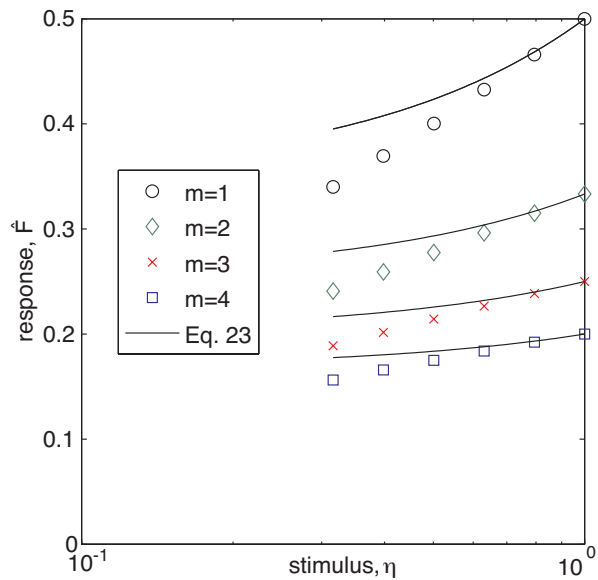, clip =, width=1.0\linewidth}
\caption{Simulation data (symbols) compare reasonably with the prediction of the perturbative approximation close to saturation, Eq. \ref{dFdeta}, for different refractory states. $\delta \eta$ was chosen to be the different between $\eta=10^{0}$ and $\eta=10^{-0.1}$, corresponding to the two rightmost data points of each simulation.
}\label{topSlope}
\end{figure}

Figure~\ref{topSlope} shows the response $\hat F$ close to $\eta = 1$ calculated for various values of $m$ from the simulation (symbols), and from Eq.~(\ref{dFdeta}) (solid lines). Eq.~(\ref{dFdeta}) describes well the slope of $\hat F$ close to $\eta = 1$. An important observation is that as $m$ grows, the relative slope $\hat F^{-1}d\hat F/d\eta$ at $\eta = 1$  decreases. Therefore, if the typical refractory period $m$ is large, the response $\hat F$ saturates [e.g., reaching $90\%$ of $\hat F(1)$] for smaller values of $\eta.$

\begin{figure}[t]
\centering
\epsfig{ file=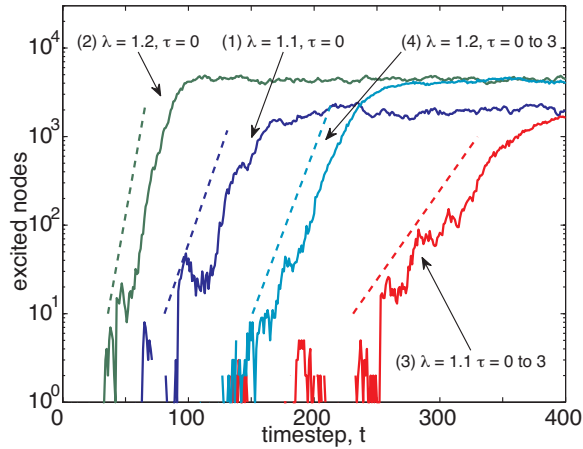, clip=,width=1\linewidth}
\caption{Time series (solid lines) for initial growth of signals within four active networks, with growth rates from Eq.~(\ref{dFdeta}) shown (dotted lines). The timesteps shown on the horizontal axis have been shifted to display multiple results together, but not rescaled or distorted. In less than 100 timesteps, all networks tested exhausted the exponential growth regime. N=100000 nodes and $\eta = 10^{-6}$, for (1) $\lambda=1.1$, $\tau=0$, (2) $\lambda=1.2$, $\tau=0$, (3) $\lambda = 1.1$, $\tau \in \{0,1,2,3\}$, (4) $\lambda=1.2$, $\tau \in \{0,1,2,3\}$.}\label{growthRateFigure}
\end{figure}

Transmission delays, as in the analogous system of gene regulatory networks\cite{genes}, do not modify steady state response. However, delays modify the time scale of relaxation to steady state. We quantified this modification in the growth rate in Eq.~(\ref{newgrowthrate}), which determines the growth rate of perturbations from an almost critical quiescent network in terms of a matrix determined from the distribution of delays. 
In Fig.~\ref{growthRateFigure} we show time series (solid lines) for the initial growth in the number of excited nodes within four active networks with and without time delays. For comparison, we show the slope that results from the corresponding growth rates obtained from Eq.~(\ref{newgrowthrate}) (dotted lines). The timesteps shown on the horizontal axis have been shifted to display multiple results together, but not rescaled or distorted.
As shown in Fig.~\ref{growthRateFigure}, Eq.~(\ref{newgrowthrate}) is helpful in quantifying the growth rate of signals within the network in the regime during which growth is exponential. In this limited regime, simulation data compare well with time series of excitations, and capture the growth rate's dependence on eigenvalue and time delays. We note here that Eq.~(\ref{newgrowthrate}) predicts the growth rate of $p_i^t$, and therefore the growth rate of both $f^t$ and $\hat f^t$. Here we have chosen to show the growth in the number of excited nodes (proportional to $f^t$) which is more experimentally accessible than $\hat f^t$.

\subsection{Validity of the approximation $Ap \propto u$}\label{validity}
Here we will address the question of the validity of our approximation $Ap \propto u$, which was used to develop the nonperturbative approximation Eq.~(\ref{nonpe}). First, we note that when $\eta$ and $p$ are small, the linear analysis of Sec.~\ref{dynamics} and Ref.\cite{Larremore} shows that $p \propto u$, and therefore the approximation $Ap \propto u$ is justified in this regime. As $\eta$ grows, and for situations where $p$ is not small, one should expect deviations of $p$ from being parallel to $u$. However, we note that since $p_i$ measures how active node $i$ is, it should still be highly correlated with the in-degree of node $i$. Since in many situations the in-degree is also correlated with the entries of the eigenvector $u$, we expect that in those cases $p$ remains correlated with $u$. After multiplication by $A$, the approximation can only become better. For the class of networks in which the ratio between the largest eigenvalue $\lambda$ and the next largest eigenvalue scales as $\sqrt{\langle d\rangle}$ (which include Erd\H{o}s-R\'{e}nyi and other networks\cite{chauhanOtt}), we expect that $Ap \propto u$ should be a good approximation.

Another reason why the approximation $Ap \propto u$ works well even when $\hat F$ is not small is that the errors introduced by this approximation vanish exactly when $\eta = 1$. To see this, note that for $\eta = 1$, since each node cycles deterministically through its $m_i+1$ available states, we have $p_i = 1/(1+m_i)$, which gives $\hat F = \sum_{i,j} A_{i,j}(1+m_j)^{-1} / \sum_{i,j} A_{i,j} = \langle d^{out} (1 + m)^{-1} \rangle/{\langle d\rangle}$, which agrees exactly with the result of setting $\eta = 1$ in Eq.~(\ref{nonpe}). Thus, even as the assumption $Ap \propto u$ may become less accurate as $\eta$ grows, the importance of the error introduced by it decreases and eventually vanishes at $\eta = 1$.

\begin{figure}[t]
\centering 
\epsfig{file =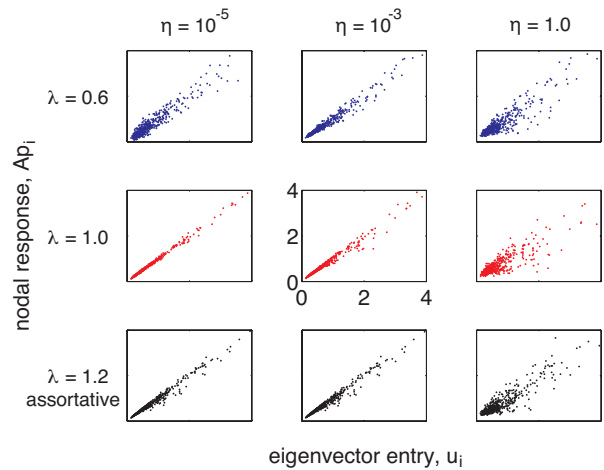, clip =,width=1\linewidth }
\caption{Plots of normalized $Ap_i$ vs $su_i$ for scale-free networks, with eigenvalues 0.6 (blue, top row), 1.0 (red, middle row), and 1.2 with assortative mixing (black, bottom row) at stimulus levels $\eta = 10^{-5}, 10^{-3}, $and 1 for the left, middle, and center columns respectively. Agreement is very good for critical and active cases, with more noise in the quiescent case due to less incoming stimuli over the duration of the simulation. 
}\label{ApEqualssu}
\end{figure}

To illustrate how the assumption $Ap \approx su$ works in some particular examples, Fig.~\ref{ApEqualssu} compares normalized $Ap_i$ and $u_i$ for a variety of eigenvalues and stimulus levels. Good agreement between them (characterized by a high correlation) indicates that the assumption of section \ref{nonperturbative} is valid, whereas more noisy agreement for some cases indicates that the assumption $Ap\propto u$ is invalid (although, as discussed above, this does not necessarily imply that the nonperturbative approach will fail). Low stimulus levels in quiescent networks (top left panel) show relatively low correlation for short simulations, but the correlation improves with more timesteps as relative nodal response increases at well connected nodes and decreases at poorly connected nodes. Assortative networks (bottom panels) show slightly lower correlation as well, corroborating the results shown in Fig.~\ref{semilogzoo} where the predictive power of Eq.~(\ref{nonpe}) is slightly diminished for the assortative network. As expected, correlation between $Ap$ and $u$ entries is worst at $\eta = 1$ (right panels), but we reiterate that for $\eta = 1$ this error does not affect the predictions of Eq.~(\ref{nonpe}).

\section{Discussion}\label{discussion}

In this paper we studied a generalized version of the Kinouchi-Copelli model in complex networks. We developed a nonperturbative treatment [Eq.(\ref{nonpe})] that allows us to find the response $\hat{F}$ of the network for a given value of the stimulus given a matrix of excitation transmission probabilities $A$. Our approach includes the possibility of heterogeneous distributions of excitation transmission delays and numbers of refractory states. An important assumption in our theory is that there are many incoming links to every node, which allows us to transform the product in Eq.~(\ref{nasty}) into an exponential. This assumption is very reasonable for neural networks, where the number of synapses per neuron is estimated\cite{neuralConnectivityBook} to be of the order of $10,000$. In addition, in order to obtain a closed equation for $\hat{F}$, we assumed $Ap \propto u$. As discussed in Sec.~\ref{validity}, this approximation works well in the regime when the response and stimulus are small. Furthermore, the error introduced by this approximation becomes smaller as the probability of stimulus increases and eventually vanishes for $\eta=1$. The result is that Eq.~(\ref{nonpe}) predicts the response $\hat F$ satisfactorily for all values of $\eta$. While we validated our predictions using scale-free networks with various correlation properties, we did not test them in topologies in which mean-field theories have been shown to fail, such as periodic hypercubes and branching tree networks \cite{assisCopelli,dendriticTrees}. This study is left to future research.

Our theory describes how the introduction of additional refractory states modifies the network response by modifying Eq.~(\ref{nonpe}). In addition, their effect is captured by the perturbative approximations of Sec.~\ref{perturbative} which, although valid in principle only for very small $\hat F$, we have found successfully predict the effect of a distribution in the number of refractory states for a larger range of response values. 

We studied the effect of time delays on the time scale needed to reach a steady-state response, and found that Eq.~(\ref{newgrowthrate}) determines the growth rate of perturbations from a quiescent, almost critical network. The temporal characteristics of the response could be important in the study of sensory systems, in which the stimulus level might be constantly changing in time. Additionally, delays may be important in studying the phenomenon of synchronization and propagation of wavefronts, which we do not study here. Synchronization in  epidemic models similar to the model considered here has been well-described in the absence of time delays \cite{girvanSynchronization}, and synchronization in Rulkov neurons has been shown to be affected subtly by time delays \cite{rulkovDelays}. However, the effect of time delays on synchronization in our model remains an open line of inquiry.

An important practical question regarding the application of our theory to neuroscience is how our results can be made compatible with the presence of excitatory and inhibitory connections in neural networks. Considering one excited neuron, and after excitatory and inhibitory connections are taken into account, the important quantity that determines the future activity of the network is how many other neurons are expected to be excited by the originally excited neuron. This number might depend on the overall balance of excitatory and inhibitory connections, but it must be a positive number. The Kinouchi-Copelli model we are using, and similar models used successfully by neuroscientists to model neuronal avalanches\cite{woody}, have therefore considered only excitatory neurons, while adjusting the probabilities of excitation transmission to account for different balances of excitatory and inhibitory neurons. Nevertheless, we believe a generalization of the Kinouchi-Copelli model that accounts for inhibitory connections should be investigated in the future.

Another important issue is the generality of our findings for more biologically realistic excitable systems. We conjecture that the effect of network topology on the dynamic range of networks of continuous-time, continuous-state coupled excitable systems such as coupled ODE neuron models\cite{izhikevich} is qualitatively similar to its effect on the class of discrete-time, discrete-state dynamical systems studied here. However, this remains open to investigation.


\section*{Acknowledgements}

We thank Dietmar Plenz for useful discussions. The work of Daniel Larremore was supported by the NSF's Mentoring Through Critical Transition Points (MCTP) Grant, No DMS-0602284. The work of Woodrow Shew was supported by the Intramural Research Program of the National Institute of Mental Health. The work of E. Ott was supported by ONR Grant N00014-07-1-0734.



\begin{thebibliography}{99}
\bibitem{endnote1} This is almost certainly not the method by which real biological networks tune their connectivity, but provides a theoretical starting point by which we can assess the properties of networks with varying $\lambda$, without commenting on how real networks may achieve such connectivities.
\end{thebibliography}
\end{document}